\documentclass[12pt,english]{article}
\usepackage[T1]{fontenc}
\usepackage[latin1]{inputenc}
\usepackage{amsmath}
\usepackage{amssymb}

\makeatletter
\usepackage{setspace}

\makeatletter
\usepackage{latexsym}

\makeatother

\usepackage{babel}
\makeatother
\begin{document}

\title{Stability of a Noncommutative Jackiw-Teitelboim Gravity}

\author{D. V. Vassilevich%
\thanks{\textit{Institut für Theoretische Physik, Universität Leipzig,} \textit{Postfach
100 920, D-04009 Leipzig, Germany}, \emph{also at V.A.Fock Institute
of Physics, St.Petersburg University, Russia: Dmitri.Vassilevich@itp.uni-leipzig.de}%
},\\
  R. Fresneda%
\thanks{\emph{Instituto de Física, Unversidade de São Paulo}, \emph{Brasil:
fresneda@fma.if.usp.br}%
}, D. M. Gitman%
\thanks{\emph{Instituto de Física, Unversidade de São Paulo}, \emph{Brasil:
gitman@dfn.if.usp.br}%
}\\
 }

\maketitle
\begin{abstract}
We start with a noncommutative version of the Jackiw-Teitelboim gravity
in two dimensions which has a linear potential for the dilaton fields.
We study whether it is possible to deform this model by adding quadratic
terms to the potential but preserving the number of gauge symmetries.
We find that no such deformation exists (provided one does not twist
the gauge symmetries). 
\end{abstract}

\section{Introduction}

Dilaton gravities in two dimensions \cite{Grumiller:2002nm} are a
good testing ground for many theoretical ideas also relevant in higher
dimensions. After some field redefinitions almost all interesting
models of that type can be written in the form \begin{equation}
S=\int d^{2}x\varepsilon^{\mu\nu}\left(\phi\partial_{\mu}\omega_{\nu}+\phi_{a}D_{\mu}e_{\nu}^{a}-\varepsilon_{ab}e_{\mu}^{a}e_{\nu}^{a}V(\phi)\right),\label{dilact}\end{equation}
 where $e_{\mu}^{a}$ is the zweibein, $\varepsilon^{\mu\nu}$ is
the Levi-Civita symbol (see Appendix \ref{appnot} for our sign conventions).
The covariant derivative \begin{equation}
\varepsilon^{\mu\nu}D_{\mu}e_{\nu}^{a}=\varepsilon^{\mu\nu}\left(\partial_{\mu}e_{\nu}^{a}+\omega_{\mu}\varepsilon_{\,\, b}^{a}e_{\nu}^{b}\right)\label{covdir}\end{equation}
 contains the spin connection $\omega_{\mu}\varepsilon_{\,\, b}^{a}$.
Here $\phi$ is a scalar field called the dilaton. $\phi_{a}$ is
an auxiliary field. In the commutative case, which we are considering
at the moment, \emph{any} choice of the potential $V(\phi)$ leads
to a consistent model. Two examples are of particular importance for
us. A constant potential $V$ corresponds to the (conformally transformed)
string gravity, also called the Witten black hole \cite{stringgrav}.
For a linear potential $V(\phi)\propto\phi$ one gets the Jackiw-Teitelboim
(JT) model \cite{JT}, whose equations of motion were studied earlier
in \cite{BaChNe}.

The auxiliary field $\phi_{a}$ generates the condition that $\omega_{\mu}$
is the Levi-Civita connection compatible with $e_{\mu}^{a}$. Under
this condition $\varepsilon^{\mu\nu}\partial_{\mu}\omega_{\nu}$ becomes
proportional to the usual Riemann curvature (the terms proportional
to $\phi_{a}$, of course, disappears). In this way one arrives at
a second order formalism, which may be more familiar to some of the
readers. However, the first order action (\ref{dilact}) has many
advantages over the second order one. For instance, the classical
equations of motion are much easier to solve \cite{Katanaev:1995bh},
and in the quantum case, it is possible to perform the path integral
over the geometric variables even in the presence of additional matter
fields \cite{2Dquant}.

In this paper we study which models of 2D dilaton gravity can be formulated
on noncommutative spaces. Let us define the star product of functions
which will replace the usual pointwise multiplication. The Moyal star
product of functions on $\mathbb{R}^{2}$ reads \begin{equation}
f\star g=f(x)\exp\left(\frac{i}{2}\,\theta^{\mu\nu}\overleftarrow{\partial}_{\mu}\overrightarrow{\partial}_{\nu}\right)g(x)\,.\label{starprod}\end{equation}
 $\theta$ is a constant antisymmetric matrix. This product is associative,
$(f\star g)\star h=f\star(g\star h)$. In this form the star product
has to be applied to plane waves and then extended to all (square
integrable) functions by means of the Fourier series. Obviously, \begin{equation}
x^{\mu}\star x^{\nu}-x^{\nu}\star x^{\mu}=i\theta^{\mu\nu}\,.\label{xmxn}\end{equation}

Furthermore, the Moyal product is closed, \begin{equation}
\int_{\mathcal{M}}d^{2}xf\star g=\int_{\mathcal{M}}d^{2}xf\times g\label{clo}\end{equation}
 (where $\times$ denotes usual pointwise product), it respects the
Leibniz rule \begin{equation}
\partial_{\mu}(f\star g)=(\partial_{\mu}f)\star g+f\star(\partial_{\mu}g),\label{Leib}\end{equation}
 and allows to make cyclic permutations under the integral \begin{equation}
\int_{\mathcal{M}}d^{2}xf\star g\star h=\int_{\mathcal{M}}d^{2}xh\star f\star g\,.\label{cyper}\end{equation}
 The complex conjugation reverses the order of factors, \begin{equation}
(f\star g)^{*}=g^{*}\star f^{*}.\label{comcon}\end{equation}
 The product (\ref{starprod}) is not the only possible choice of
an associative noncommutative product. The right hand side of (\ref{xmxn})
can depend, in principle, on the coordinates.

An important step towards constructing a satisfactory noncommutative
gravity was recently made by Wess and collaborators \cite{Aschieri:2005yw},
who understood how one can construct diffeomorphism invariants, including
the Einstein-Hilbert action, on noncommutative spaces (see also \cite{Zupnik:2005ph}
for a real formulation). There is, however, a price to pay. The diffeomorphism
group becomes twisted, there is a non-trivial coproduct due to which
the action of the symmetries on tensor products looks very unusual
\cite{Chaichian:2004za,Wess:2003da}.

In two dimensions it is possible to construct noncommutative (dilaton)
gravity models with an usual (non-twisted) realization of gauge symmetries.
A noncommutative version of the Jackiw-Teitelboim (NCJT) model was
constructed in \cite{Cacciatori:2002ib} and then quantised in \cite{Vassilevich:2004ym}.
A noncommutative Witten black hole model was suggested in \cite{Vassilevich:2005fk}.
Both these models are of the Yang-Mills type: the JT model is equivalent
to a topological BF model; the Witten black hole may be represented
as a Wess-Zumino-Novikov-Witten model. There are some general procedures
of how such models can be formulated in the noncommutative case (see
\cite{Cacciatori:2002ib,Ghezelbash:2000pz}). It is important therefore
to check whether one can go beyond the Yang-Mills paradigm. Besides,
if we are on the right track, dilaton gravities should exist not only
for linear or constant potentials, but also for an arbitrary potential
$V$. In the present paper we study whether quadratic potentials are
allowed.

To analyze the gauge symmetries we use the canonical formalism for
noncommutative spacetime developed in \cite{Vassilevich:2005fk}.
This is not a canonical formalism in the usual sense of the word%
\footnote{Since the space-time noncommutative theories are non-local in time
and contain an infinite number of time derivatives hidden in the star
product, it is obvious that some modification of the standard canonical
formalism is necessary.%
} \cite{Gitman:1990qh,Henneaux:1992ig}, but it makes it possible to
define the notion of first class constraints and to associate a gauge
symmetry to them. As to commutative gauge theories, it was conjectured
by Dirac that all first-class constraints act as generators of gauge
transformations. For some classes of commutative gauge theories this
conjecture can be proved and, in addition, it turns out that the number
of independent non-trivial gauge transformations is equal to the number
of primary first-class constraints \cite{Gitman:1990qh}. The symmetry
structure of a general commutative gauge theory was recently described
in detail and related to the constraint structure of the theory in
the Hamiltonian formulation \cite{Gitman:2004zf}. In particular,
the gauge charge was constructed explicitly as a decomposition in
the special orthogonal constraint basis. It was demonstrated that,
in the general case, the gauge charge cannot be constructed with the
help of first-class constraints alone, for its decomposition also
contains special combinations of second-class constraints.

Consider those classical actions which can be represented in the form
\begin{equation}
S=\int d^{2}x\left(p^{i}\partial_{0}q_{i}-\lambda^{i}\star G_{i}(p,q)\right)\,,\label{genact}\end{equation}
 so that the expressions ('constraints') $G_{i}(p,q)$ do not contain
\emph{explicit} time derivatives (implicit time derivatives are always
present through the star product). The paper \cite{Vassilevich:2005fk}
demonstrated that one can define the canonical pairs ignoring implicit
time derivatives in the star product. In this sense $p^{i}$ become
canonically conjugated to $q_{i}$. The brackets are then defined
by the equation \begin{equation}
\{ q_{i}(x),p^{j}(y)\}=\delta_{i}^{j}\delta^{2}(x-y).\label{qipj}\end{equation}
 This definition can be extended to all polynomial functionals on
the phase space (see \cite{Vassilevich:2005fk}). If the brackets
between the constraints are again linear combinations of constraints,
then the noncommutative action has a gauge symmetry associated to
each $G_{i}$. In this sense, the $G_{i}$ may be called first class
constraints.

The most unusual property of the bracket (\ref{qipj}) is the presence
of the delta-function of the time coordinates on the right hand side.
However, since the space-time noncommutative theories are nonlocal
in the time direction, restriction of the brackets of the phase space
variables calculated at the same value of time does not look natural
and even consistent. The presence of an additional delta-function
in (\ref{qipj}) reminds us of the Ostrogradski formalism for the
theories with higher temporal derivatives (see \cite{Ostro,key-3}
and \cite{Vassilevich:2005fk} for a more extensive discussion). Anyway,
one can also use the brackets (\ref{qipj}) to analyse gauge symmetries
in commutative theories. It is not clear, however, whether one can
use the modified brackets for quantization. In the present paper we
shall exclusively use (\ref{qipj}) to define the Poisson structure.

We shall demonstrate that one cannot consistently add quadratic terms
to the dilaton potential of the NCJT model, so that it is \emph{stable}
against such deformations.

\section{Noncommutative Jackiw-Teitelboim gravity}

A noncommutative version of the Jackiw-Teitelboim model has been constructed
in \cite{Cacciatori:2002ib}. It has been identified with a $U(1,1)$
gauge theory on noncommutative $\mathbb{R}^{2}$. The action reads
\begin{equation}
S^{(0)}=\frac{1}{4}\int d^{2}x\,\varepsilon^{\mu\nu}\left[\phi_{ab}\star\left(R_{\mu\nu}^{ab}-2\Lambda e_{\mu}^{a}\star e_{\nu}^{b}\right)-2\phi_{a}\star T_{\mu\nu}^{a}\right]\label{actJT}\end{equation}
 with curvature tensor \begin{align}
R_{\mu\nu}^{ab}= & \varepsilon^{ab}\left(\partial_{\mu}\omega_{\nu}-\partial_{\nu}\omega_{\mu}+\frac{i}{2}[\omega_{\mu},b_{\nu}]+\frac{i}{2}[b_{\mu},\omega_{\nu}]\right)\nonumber \\
 & +\eta^{ab}\left(i\partial_{\mu}b_{\nu}-i\partial_{\nu}b_{\mu}+\frac{1}{2}[\omega_{\mu},\omega_{\nu}]-\frac{1}{2}[b_{\mu},b_{\nu}]\right)\label{Rtensor}\end{align}
 and with noncommutative torsion \begin{align}
T_{\mu\nu}^{a}=\partial_{\mu}e_{\nu}^{a}-\partial_{\nu}e_{\mu}^{a} & +\frac{1}{2}{\varepsilon^{a}}_{b}\left([\omega_{\mu},e_{\nu}^{b}]_{+}-[\omega_{\nu},e_{\mu}^{b}]_{+}\right)\nonumber \\
 & +\frac{i}{2}\left([b_{\mu},e_{\nu}^{a}]-[b_{\nu},e_{\mu}^{a}]\right)\,.\label{torsion}\end{align}
 There are two dilaton fields, $\phi$ and $\psi$, which are combined
into \begin{equation}
\phi_{ab}:=\phi\varepsilon_{ab}-i\eta_{ab}\psi\,.\label{pab}\end{equation}
 All commutators (denoted by square brackets) and anticommutators
(denoted by $[\,\,,\,\,]_{+}$) are calculated with the Moyal star
product.

We note that (\ref{actJT}) contains more fields than the initial
commutative model. This is related to the fact that the gauge group
of the commutative JT model, which is $SU(1,1)$, cannot be closed
on the noncommutative plane. To make the closure, one introduces additional
$U(1)$ fields $\psi$ and $b_{\mu}$ which decouple in the commutative
limit.

One can rewrite (\ref{actJT}) in the canonical form: \begin{equation}
S^{(0)}=\int d^{2}x\left(p^{i}\partial_{0}q_{i}-\lambda^{i}\star G_{i}^{(0)}\right)\,,\label{canact}\end{equation}
 where \begin{align}
 & q_{i}=(e_{1}^{a},\omega_{1},b_{1}),\nonumber \\
 & p^{i}=(\phi_{a},\phi,-\psi),\label{qpl}\\
 & \lambda^{i}=(e_{0}^{a},\omega_{0},b_{0}).\nonumber \end{align}
 The constraints are \begin{align}
 & G_{a}^{(0)}=-\partial_{1}\phi_{a}+\frac{1}{2}{\varepsilon^{b}}_{a}[\omega_{1},\phi_{b}]_{+}+\frac{i}{2}[\phi_{a},b_{1}]\nonumber \\
 & \qquad\qquad\qquad+\frac{\Lambda}{2}\left(-\varepsilon_{ab}[e_{1}^{b},\phi]_{+}+i\eta_{ab}[e_{1}^{b},\psi]\right)\,,\label{Ga}\\
 & G_{3}^{(0)}=-\partial_{1}\phi+\frac{i}{2}[\phi,b_{1}]+\frac{i}{2}[\psi,\omega_{1}]-\frac{1}{2}{\varepsilon^{a}}_{b}[\phi_{a},e_{1}^{b}]_{+}\,,\label{G3}\\
 & G_{4}^{(0)}=\partial_{1}\psi-\frac{i}{2}[\psi,b_{1}]+\frac{i}{2}[\phi,\omega_{1}]+\frac{i}{2}[\phi_{a},e_{1}^{a}]\,.\label{G4}\end{align}

It is easy to check that the constraint algebra closes, and the brackets
between the constraints read \begin{eqnarray}
 &  & \left\{ \int\alpha^{a}\star G_{a}^{(0)},\int\beta^{b}\star G_{b}^{(0)}\right\} =\nonumber \\
 &  & \qquad\qquad\qquad=-\frac{\Lambda}{2}\int\left(\varepsilon_{ab}[\alpha^{a},\beta^{b}]_{+}\star G_{3}^{(0)}+i[\alpha_{a},\beta^{a}]\star G_{4}^{(0)}\right)\label{0ab}\\
 &  & \left\{ \int\alpha\star G_{3}^{(0)},\int\beta\star G_{3}^{(0)}\right\} =\frac{i}{2}\int[\alpha,\beta]\star G_{4}^{(0)}\label{033}\\
 &  & \left\{ \int\alpha\star G_{4}^{(0)},\int\beta\star G_{4}^{(0)}\right\} =-\frac{i}{2}\int[\alpha,\beta]\star G_{4}^{(0)}\label{044}\\
 &  & \left\{ \int\alpha\star G_{3}^{(0)},\int\beta\star G_{4}^{(0)}\right\} =-\frac{i}{2}\int[\alpha,\beta]\star G_{3}^{(0)}\label{034}\\
 &  & \left\{ \int\alpha\star G_{3}^{(0)},\int\beta^{a}\star G_{a}^{(0)}\right\} =-\frac{1}{2}\int[\alpha,\beta^{a}]_{+}\,{\varepsilon^{b}}_{a}\star G_{b}^{(0)}\label{03a}\\
 &  & \left\{ \int\alpha\star G_{4}^{(0)},\int\beta^{a}\star G_{a}^{(0)}\right\} =-\frac{i}{2}\int[\alpha,\beta^{a}]\star G_{a}^{(0)}\label{04a}\end{eqnarray}
 Here we introduced a short-hand notation $\int:=\int d^{2}x$.

\section{Deformations}

Let us now discuss deformations of the NCJT model. We shall add some
terms to the action (\ref{actJT}) so that (i) the field content of
the model will not be changed, and (ii) the number of secondary first
class constraints (and, consequently, the number of gauge symmetries)
will also remain invariant. Being inspired by commutative dilaton
gravity models we only consider the deformations of the potential
term, and we only add terms of the next (quadratic) order in the two
dilaton fields $\phi$ and $\psi$.

In addition to analogies with the commutative case, there are also
other reasons for not considering deformations of the curvature and
torsion terms. For example, replacing $\phi_{ab}$ in (\ref{pab})
by a non-linear function of the dilatons is equivalent to a redefinition
of the dilaton fields. Adding higher powers of the curvature in general
adds new degrees of freedom to the theory, and this is a more drastic
modification than it is usually understood as deformations. The same
also refers to torsion terms.

Further restrictions on possible deformations are imposed by the global
symmetries of the model which we would like to preserve. First of
all, we require the symmetry with respect to global rotation of the
tangential and world indices. This implies that all indices must be
contracted pair-wise. We also require that the terms being added are
of even parity. Since $\phi$ is a scalar, and $\psi$ is a pseudo-scalar,
even (odd) powers of $\psi$ should be multiplied with even (odd)
powers of the Levi-Civita symbol $\varepsilon$. As a result, we obtain
the following family of quadratic deformations of the NCJT model \begin{equation}
S=S^{(0)}+\tilde{S},\label{Sdef}\end{equation}
 where \begin{align}
 & \tilde{S}=\int d^{2}x\left(\varepsilon^{\mu\nu}\varepsilon_{ab}\left(c_{1}e_{\mu}^{a}\star e_{\nu}^{b}\star\phi^{2}+c_{2}e_{\mu}^{a}\star e_{\nu}^{b}\star\psi^{2}\right.\right.\nonumber \\
 & \qquad\qquad\qquad\left.+c_{3}e_{\mu}^{a}\star\phi\star e_{\nu}^{b}\star\phi+c_{4}e_{\mu}^{a}\star\psi\star e_{\nu}^{b}\star\psi\right)\nonumber \\
 & \qquad\qquad+\varepsilon^{\mu\nu}\eta_{ab}\left(c_{5}e_{\mu}^{a}\star e_{\nu}^{b}\star[\phi,\psi]+ic_{6}e_{\mu}^{a}\star e_{\nu}^{b}\star[\phi,\psi]_{+}\right.\nonumber \\
 & \qquad\qquad\qquad\left.\left.+\frac{i}{2}c_{7}(e_{\mu}^{a}\star\phi\star e_{\nu}^{b}\star\psi-e_{\mu}^{a}\star\psi\star e_{\nu}^{b}\star\phi)\right)\right)\,.\label{defterms}\end{align}
 The arbitrary constants $c_{1}$, $c_{2}$, ..., $c_{7}$ must be
real to preserve the reality of the total action $S$. The powers
are taken with the star-product, for example $\phi^{2}\equiv\phi\star\phi$.

The constraints read \begin{equation}
G_{a}=G_{a}^{(0)}+\tilde{G}_{a},\quad G_{3}=G_{3}^{(0)},\quad G_{4}=G_{4}^{(0)},\label{newcon}\end{equation}
 where \begin{align}
\tilde{G}_{a}= & \varepsilon_{ab}\left(c_{1}[e_{1}^{b},\phi^{2}]_{+}+c_{2}[e_{1}^{b},\psi^{2}]_{+}+2c_{3}\phi e_{1}^{b}\phi+2c_{4}\psi e_{1}^{b}\psi\right)\nonumber \\
 & +\eta_{ab}\left(c_{5}[e_{1}^{b},[\phi,\psi]]+ic_{6}[e_{1}^{b},[\phi,\psi]_{+}]+ic_{7}(\phi e_{1}^{b}\psi-\psi e_{1}^{b}\phi)\right).\label{tilGa}\end{align}

Our next step is to check whether the constraint algebra still closes
on the constraint surface%
\footnote{In principle, other substantial modifications of the constraint algebra
may occur, but not in the present case. We limit the number of gauge
symmetries to four, so only four first class constraints are allowed,
because there are only four canonical pairs of variables. Therefore,
the only possibility is that $G_{i}$ are first class and that their
brackets give again linear combinations of $G_{i}$.%
}. Since the constraints $G_{3}$ and $G_{4}$ are unchanged, the brackets
between them (\ref{033}) - (\ref{034}) are the same. It is an easy
exercise to check that for all values of the constants $c_{m}$\begin{equation}
\left\{ \int\alpha\star G_{4},\int\beta^{a}\star\tilde{G}_{a}\right\} =-\frac{i}{2}\int[\alpha,\beta^{a}]\star\tilde{G}_{a}\,.\label{4a}\end{equation}
 Consequently, for any values of $c_{m}$ the bracket between $G_{4}$and
$G_{a}$, \begin{equation}
\left\{ \int\alpha\star G_{4},\int\beta^{a}\star G_{a}\right\} =-\frac{i}{2}\int[\alpha,\beta^{a}]\star G_{a}\,,\label{full4a}\end{equation}
 is again a constraint in the new set (\ref{newcon}), so that we
are getting no restrictions on $c_{m}$.

Let us now consider the bracket between $G_{3}$ and $G_{a}$, \begin{align}
 & \left\{ \int\alpha\star G_{3},\int\beta^{a}\star\tilde{G}_{a}\right\} =\frac{1}{2}\int\left[c_{1}\left(\beta_{a}\star[[\alpha,e_{1}^{a}]_{+},\phi^{2}]_{+}\right.\right.\nonumber \\
 & \quad\left.+i\beta^{a}\star\varepsilon_{ab}[e_{1}^{b},[[\alpha,\psi],\phi]_{+}]_{+}\right)+c_{2}\left(\beta_{a}\star[[\alpha,e_{1}^{a}]_{+},\psi^{2}]_{+}\right.\nonumber \\
 & \quad\left.-i\beta^{a}\star\varepsilon_{ab}[e_{1}^{b},[[\alpha,\phi],\psi]_{+}]_{+}\right)+2c_{3}\left(\beta_{a}\star\phi\star[\alpha,e_{1}^{a}]_{+}\star\phi\right.\nonumber \\
 & \quad\left.+i\beta^{a}\star\varepsilon_{ab}([\alpha,\psi]\star e_{1}^{b}\star\phi+\phi\star e_{1}^{b}\star[\alpha,\psi])\right)+2c_{4}\left(\beta_{a}\star\psi\star[\alpha,e_{1}^{a}]_{+}\star\psi\right.\nonumber \\
 & \quad\left.-i\beta^{a}\star\varepsilon_{ab}([\alpha,\phi]\star e_{1}^{b}\star\psi+\psi\star e_{1}^{b}\star[\alpha,\phi])\right)\nonumber \\
 & \quad+c_{5}\left(\beta^{a}\star\varepsilon_{ab}[[\alpha,e_{1}^{b}]_{+},[\phi,\psi]]+i\beta_{a}\star[e_{1}^{a},[[\alpha,\psi],\psi]-[\phi,[\alpha,\phi]]]\right)\nonumber \\
 & \quad+ic_{6}\left(\beta^{a}\star\varepsilon_{ab}[[\alpha,e_{1}^{b}]_{+},[\phi,\psi]_{+}]+i\beta_{a}\star[e_{1}^{a},[[\alpha,\psi],\psi]_{+}-[\phi,[\alpha,\phi]]_{+}]\right)\nonumber \\
 & \quad+ic_{7}\left(\beta^{a}\star\varepsilon_{ab}(\phi\star[\alpha,e_{1}^{b}]_{+}\star\psi-\psi\star[\alpha,e_{1}^{b}]_{+}\star\phi)\right.\label{G3tGa}\\
 & \quad\left.\left.+i\beta_{a}\star([\alpha,\psi]\star e_{1}^{a}\star\psi-\phi\star e_{1}^{a}\star[\alpha,\phi]+[\alpha,\phi]\star e_{1}^{a}\star\phi-\psi\star e_{1}^{a}\star[\alpha,\psi])\right)\right]\,.\nonumber \end{align}
 First we observe that the right hand side of (\ref{G3tGa}) contains
no terms with derivatives. This excludes the possibility of the bracket
(\ref{G3tGa}) containing any terms proportional to (\ref{Ga}), (\ref{G3}),
or (\ref{G4}). Therefore, this bracket can only be proportional to
(\ref{tilGa}), with coefficients (structure functions) as in (\ref{03a}),
so that the bracket between $G_{3}$ and $G_{a}$ sums up to become
\begin{equation}
\left\{ \int\alpha\star G_{3},\int\beta^{a}\star G_{a}\right\} =-\frac{1}{2}\int[\alpha,\beta^{a}]_{+}\,{\varepsilon^{b}}_{a}\star G_{b}\,.\label{nG3Ga}\end{equation}
 We have to compare the expressions on both sides of (\ref{nG3Ga})
to get restrictions on the constants $c_{m}$. There are no monomials
on the right hand side of (\ref{nG3Ga}) which are second order in
$\phi$ and have an explicit $i$ factor. At the same time, there
is such a term proportional to $c_{5}$ in (\ref{G3tGa}). Since all
$c_{m}$ are real, we conclude that \begin{equation}
c_{5}=0.\label{c5}\end{equation}
 Next we compare the terms in which two $\phi$ appear next to each
other%
\footnote{This also includes the terms which can be put in this form by using
property (\ref{cyper}).%
} (combined in $\phi^{2}$). Those terms agree on both sides of (\ref{nG3Ga})
if and only if \begin{equation}
c_{6}=-c_{1}.\label{c6c1}\end{equation}
 By comparing the terms where two fields $\phi$ appear separated
by other fields, we obtain the following condition \begin{equation}
2c_{3}=-c_{7}.\label{c3c7}\end{equation}
 Then we repeat the same procedure with the terms which are quadratic
in $\psi$ to get \begin{equation}
c_{2}=c_{6},\qquad2c_{4}=-c_{7}.\label{c2647}\end{equation}
 The comparison of mixed terms (containing both $\phi$ and $\psi$)
does not produce any additional restrictions on $c_{m}$. We conclude
that only two independent constants (say, $c_{1}$ and $c_{7}$) remain,
so $\tilde{G}_{a}$ can be rewritten as \begin{align}
\tilde{G}_{a}= & c_{1}\left(\varepsilon_{ab}[e_{1}^{b},\phi^{2}-\psi^{2}]_{+}-i\eta_{ab}[e_{1}^{b},[\phi,\psi]_{+}]\right)\nonumber \\
 & +c_{7}\left(-\varepsilon_{ab}(\phi\star e_{1}^{b}\star\phi+\psi\star e_{1}^{b}\star\psi)+i\eta_{ab}(\phi\star e_{1}^{b}\star\psi-\psi\star e_{1}^{b}\star\phi)\right).\label{newtGa}\end{align}

It remains to study the brackets between $G_{a}$ and $G_{b}$. Obviously,
the brackets between $\tilde{G}_{a}$ and $\tilde{G}_{b}$ vanish,
so that all new information is contained in the brackets between $G_{a}^{(0)}$
and $\tilde{G}_{b}$. The strategy is the same as above. First we
analyze the derivative terms \begin{eqnarray}
 &  & \left\{ \int\alpha^{a}\star G_{a}^{(0)},\int\beta^{b}\star\tilde{G}_{b}\right\} +\left\{ \int\alpha^{a}\star\tilde{G}_{a},\int\beta^{b}\star G_{b}^{(0)}\right\} =\label{1652}\\
 &  & =\int\left[c_{1}\left(\partial_{1}\phi\star([\phi,\varepsilon_{bc}[\beta^{b},\alpha^{c}]_{+}]_{+}+i[\psi,[\alpha_{b},\beta^{b}]]_{+})\right.\right.\nonumber \\
 &  & \qquad+\left.\partial_{1}\psi\star(-[\psi,\varepsilon_{bc}[\beta^{b},\alpha^{c}]_{+}]_{+}+i[\phi,[\alpha_{b},\beta^{b}]]_{+})\right)\nonumber \\
 &  & \quad+c_{7}\left(\partial_{1}\phi\star(-\varepsilon_{bc}(\beta^{b}\star\phi\star\alpha^{c}+\alpha^{c}\star\phi\star\beta^{b})+i(\alpha^{b}\star\psi\star\beta_{b}-\beta_{b}\star\psi\star\alpha^{b}))\right.\nonumber \\
 &  & \qquad\left.\left.-\partial_{1}\psi\star(\varepsilon_{bc}(\beta^{b}\star\psi\star\alpha^{c}+\alpha^{c}\star\psi\star\beta^{b})+i(\alpha^{b}\star\phi\star\beta_{b}-\beta_{b}\star\phi\star\alpha^{b}))\right)\right]\nonumber \\
 &  & \quad+\mbox{non-derivative terms.}\nonumber \end{eqnarray}
 From this equation we see that, since the bracket between $G_{a}$
and $G_{b}$ must be a linear combination of the constraints (\ref{newcon}),
the constraints appearing on the right hand side can only be $G_{3}$
and $G_{4}$, since the derivative $\partial_{1}\phi_{a}$ belonging
to $G_{a}$ is not present. In fact, one can also obtain the structure
functions from (\ref{1652}), but their precise form will not be needed.
Let us consider the terms in the bracket which contain the zweibein
$e_{1}^{a}$ and the dilaton $\phi$. \begin{align}
 & \left\{ \int\alpha^{a}\star G_{a}^{(0)},\int\beta^{b}\star\tilde{G}_{b}\right\} =\nonumber \\
 & =\int\left[\frac{c_{1}}{2}\left(\varepsilon_{bc}[\beta^{b},e_{1}^{c}]_{+}\star[\phi,\varepsilon_{\,\, a}^{d}[\alpha^{a},\phi_{d}]_{+}]_{+}-[\beta_{b},e_{1}^{b}]\star[\phi,[\alpha^{a},\phi_{a}]]_{+}\right)\right.\nonumber \\
 & \qquad+\frac{c_{7}}{2}\left(-\varepsilon_{bc}\beta^{b}\star\varepsilon_{\,\, a}^{d}([\alpha^{a},\phi_{d}]_{+}\star e_{1}^{c}\star\phi+\phi\star e_{1}^{c}\star[\alpha^{a},\phi_{d}]_{+})\right.\nonumber \\
 & \qquad\quad+\left.\left.\beta_{b}(\phi\star e_{1}^{b}\star[\alpha^{a},\phi_{a}]-[\alpha^{a},\phi_{a}]\star e_{1}^{b}\star\phi)\right)\right]\label{1741}\\
 & \qquad+\mbox{terms without $e_{1}^{b}$ or $\phi$.}\nonumber \end{align}
 The arguments presented above show that if the bracket (\ref{1652})
closes on existing constraints, these constraints are $G_{3}$ and
$G_{4}$, and the structure functions depend on $\phi$ and $\psi$.
In both $G_{3}$ and $G_{4}$ the fields $e_{1}^{a}$ and $\phi_{b}$
appear in the combinations $[\phi_{a},e_{1}^{b}]$ or $[\phi_{a},e_{1}^{b}]_{+}$,
i.e. they stay next to each other. Therefore, all terms where $\phi_{b}$
and $e_{1}^{a}$ appear separated by other fields should vanish. Let
us check whether this can be achieved by adjusting the remaining parameters
$c_{1}$ and $c_{7}$. Let us study the terms with $\phi$, $\phi_{0}$,
$\alpha^{0}$, $\beta^{0}$, $e_{1}^{0}$ where $\alpha^{0}$ and
$\beta^{0}$ stay next to each other, but $\phi_{0}$ and $e_{1}^{0}$
are separated. All such terms in (\ref{1652}) can be easily collected
with the help of (\ref{1741}). They read \begin{equation}
\int\frac{c_{1}}{2}[\alpha^{0},\beta^{0}]\star(\phi_{0}\star\phi\star e_{1}^{0}-e_{1}^{0}\star\phi\star\phi_{0}).\label{1657}\end{equation}
 Since they are not allowed we conclude \begin{equation}
c_{1}=0.\label{c1}\end{equation}
 Let us now collect all other terms with the same field components
where again $\phi_{0}$ and $e_{1}^{0}$ are separated but without
any restrictions on the placement of $\alpha^{0}$ and $\beta^{0}$.
\begin{equation}
\int\frac{c_{7}}{2}[e_{1}^{0},\phi]\star(\beta^{0}\star\phi_{0}\star\alpha^{0}-\alpha^{0}\star\phi_{0}\star\beta^{0}).\label{1707}\end{equation}
 Such terms are also not allowed. Therefore, \begin{equation}
c_{7}=0.\label{c7}\end{equation}
 We have just demonstrated that no consistent quadratic deformation
of the NCJT model exists. This means that the NCJT model is stable
against such deformations.

\section{Conclusions}

In this paper we studied whether it is possible to deform the action
of the NCJT model by adding quadratic terms to the dilaton potential
while preserving the number of first-class constraints. The answer
we obtained is negative. This, of course, does not exclude the existence
of interesting NC gravity models. There is still the possibility of
existing other interacting NC dilaton gravities with usual (non-twisted)
gauge symmetries. However, it is clear that most of the commutative
dilaton gravity models (which admit arbitrary dilaton potentials)
cannot be extended to the noncommutative set-up in this approach.
Therefore, our results may be considered as a strong argument in favour
of the {}``twisted'' approach \cite{Aschieri:2005yw}, which allows
practically arbitrary self-interactions of scalar fields. We also
point out some earlier results \cite{Grumiller:2003df} which show
that deformations of 2D gravities are trivial if one does not introduce
certain amount of the quantum group structure. Another important result
is the construction of twisted conformal symmetries in two dimensions
\cite{Lizzi:2006xi}. To incorporate twisted symmetries in the canonical
formalism one should probably include twists into the canonical formalism
itself.

Finally, since the spherical reduction of higher-dimensional Einstein
gravities produces some dilaton gravities in two dimensions, one can
expect that our no-go result can be somehow extended to higher dimensions.

\section*{Acknowledgments}

This work was supported in part by the DFG project BO 1112/13-1.\\
 D.M.G. is grateful to the foundations FAPESP and CNPq for permanent
support, and R.F. would like to thank FAPESP for their financial support.

\appendix
%dummy comment inserted by tex2lyx to ensure that this paragraph is not empty

\section{Notations and useful identities}

\label{appnot} Our sign conventions are taken from \cite{Grumiller:2002nm}.
We use the tensor $\eta^{ab}=\eta_{ab}={\mathrm{diag}}(+1,-1)$ to
move indices up and down. The Levi-Civita tensor is defined by $\varepsilon^{01}=-1$,
so that the following relations hold \begin{equation}
\varepsilon^{10}=\varepsilon_{01}=1,\qquad{\varepsilon^{0}}_{1}={\varepsilon^{1}}_{0}=-{\varepsilon_{0}}^{1}=-{\varepsilon_{1}}^{0}=1\,.\label{epsrel}\end{equation}
 These relations are valid for both $\varepsilon^{ab}$ and $\varepsilon^{\mu\nu}$.
Note, that $\varepsilon^{\mu\nu}$ is always used with both indices
up.

The following useful identities hold for arbitrary functions $A_{1}$,
$A_{2}$, $B_{1}$ and $B_{2}$: \begin{eqnarray}
 &  & \int([A_{1},B_{1}]\star[B_{2},A_{2}]-[B_{1},A_{2}]\star[A_{1},B_{2}])=\nonumber \\
 &  & \qquad\qquad\qquad=-\int[A_{1},A_{2}]\star[B_{1},B_{2}]\label{com1}\\
 &  & \int([A_{1},B_{1}]_{+}\star[A_{2},B_{2}]_{+}-[A_{1},B_{2}]_{+}\star[A_{2},B_{1}]_{+})=\nonumber \\
 &  & \qquad\qquad\qquad=-\int[A_{1},A_{2}]\star[B_{1},B_{2}]\label{com2}\\
 &  & \int([A_{1},B_{1}]_{+}\star[B_{2},A_{2}]-[B_{1},A_{2}]_{+}\star[A_{1},B_{2}])=\nonumber \\
 &  & \qquad\qquad\qquad=\int[B_{1},B_{2}]\star[A_{1},A_{2}]_{+}\label{com3}\\
 &  & \int([A_{1},B_{1}]\star[A_{2},B_{2}]-[A_{1},B_{2}]_{+}\star[A_{2},B_{1}]_{+})=\nonumber \\
 &  & \qquad\qquad\qquad=-\int[A_{1},A_{2}]_{+}\star[B_{1},B_{2}]_{+}\label{com4}\end{eqnarray}

By means of the formula \begin{equation}
\varepsilon_{ab}\varepsilon_{cd}=\eta_{bc}\eta_{ad}-\eta_{ac}\eta_{bd}\label{epseps}\end{equation}
 one can get rid of repeated $\varepsilon$-symbols.

\end{document}